\documentclass[preprint,authoryear,12pt]{elsarticle}%

\usepackage{amssymb}
\usepackage{graphicx}
\usepackage{epsfig}
\usepackage{amsthm}
\usepackage{longtable}
\usepackage{lscape}
\usepackage{amsmath}

\journal{Journal of High Energy Astrophysics}

\begin{document}

\begin{frontmatter}



\title{Constraining the Mass of the Photon with Gamma-Ray Bursts}


\author[label1]{Bo Zhang}
\author[label2,label3]{Ya-Ting Chai}
\author[label2]{Yuan-Chuan Zou}
\author[label1,label4]{Xue-Feng Wu\corref{dip}}
\ead{xfwu@pmo.ac.cn}

\address[label1]{Purple Mountain Observatory, Chinese Academy of Sciences, Nanjing 210008, China}
\address[label2]{School of Physics, Huazhong University of Science and Technology, Wuhan 430074, China}
\address[label3]{Department of Physics, University of Hong Kong, \\Hong Kong, China}
\address[label4]{Joint Center for Particle, Nuclear Physics and Cosmology, Nanjing University-Purple Mountain Observatory, \\Nanjing 210008, China}
\cortext[dip]{Corresponding author.}

\begin{abstract}
One of the cornerstones of modern physics is Einstein's special relativity, with its constant speed of light and zero photon mass assumptions. Constraint on the rest mass $m_{\gamma}$ of photons is a fundamental way to test Einstein's theory, as well as other essential electromagnetic and particle theories. Since non-zero photon mass can give rise to frequency-(or energy-) dependent dispersions, measuring the time delay of photons with different frequencies emitted from explosive astrophysical events is an important and model-independent method to put such a constraint. The cosmological gamma-ray bursts (GRBs), with short time scales, high redshifts as well as broadband prompt and afterglow emissions, provide an ideal testbed for $m_{\gamma}$ constraints. In this paper we calculate the upper limits of the photon mass with GRB early time radio afterglow observations as well as multi-band radio peaks, thus improve the results of Schaefer (1999) by nearly half an order of magnitude.
\end{abstract}

\begin{keyword}
Radio continuum: general, gamma-ray burst: general, photon rest mass.
\end{keyword}

\end{frontmatter}


\def\astrobj#1{#1}
\section{Introduction}
\label{sect:intro}

Modern physics theories, especially Maxwellian electromagnetism, Einstein's theory of Special Relativity as well as quantum electrodynamics (QED) are all based on a simple precondition: the speed of light in vacuum is a constant $c$ for all electromagnetic waves, from radio to the $\gamma$-ray band. That is, photons are massless particles. Thus, whether the speed of light is a constant across the electromagnetic spectrum plays a fundamental role in physics. If photons have non-zero rest mass, various key theories should be affected. Although enormous successes have been achieved based upon the theories aforementioned, it is still necessary to put the massless photon assumption to test with as many independent methods as possible.

The mass of photon can be measured via two types of approaches, laboratory experiments and astronomical observations (Tu et al. 2005, and references therein). According to Heisenberg's uncertainty principle, currently the best limit of photon rest mass $m_{\gamma}$ one can achieve should be $m_{\gamma} \approx \hbar/c^2 \Delta T$, here $\hbar$ is the Planck constant, and $\Delta T$ the age of the Universe. With $\Delta T$ known to be in the order of $10^{10}$ years, the value of $m_{\gamma}$ inferred from various tests should be no smaller than $\approx 10^{-66}$ g (Tu et al. 2005). The laboratory measurements based upon Coulomb's law, Amp\`{e}re's law, and other electromagnetic phenomena have achieved very stringent constraints, with a lower limit of $m_{\gamma} < \left(0.7\pm 1.7\right) \times 10^{-52}$ g from measurements of torque on rotating magnetized toroid (Tu et al. 2006). On the other hand, astronomical tests utilizing various principles, from dispersions of astrophysical radiations caused by non-zero photon mass, to magneto-hydrodynamics (MHD) related phenomena, can provide independent tests on the mass limit of photons other than laboratory measurements.

Currently the mass limit of photon adopted by Particle Data Group is $m_{\gamma} \le 1.783\times 10^{-51}$g, nearly 24 orders of magnitude smaller than electron's mass $m_e$ (Olive et al. 2014), and this value is measured from observations of MHD phenomena in planetary magnetic fields (Ryutov 2007). Besides, Chibisov (1976) obtained a more stringent limit of $m_{\gamma} \le 3\times 10^{-60}$g, by analysing the stability of magnetized gas in galaxies, although this result depends on the applicability of virial theorem and other assumptions. Goldhaber \& Nieto (2003) also put a limit on the mass of photon of $10^{-52}$g considering the stability of plasma in Coma Cluster. However, since astronomical MHD phenomena can be quite complicated, such constraints could be more model-dependent in many cases.

Besides, Accioly \& Pazszko (2004) obtained a result of $m_\gamma \le 10^{-40}$g based on gravitational deflections of radio waves. While Schaefer (1999) got a limit of $m_\gamma \le 4.2 \times 10^{-44}$ by comparing the time-delay of photons with different frequencies from a wide range of explosive astronomical events, including gamma ray bursts (GRBs for short), Type Ia supernovae with high redshift, TeV flares from blazar Mrk 421, as well as the Crab Pulsar. And with a lower observational frequency as well as a shorter time scale, the time delay between the radio afterglow and prompt $\gamma$-ray emission of GRB 980703 yielded the best result in this analysis. Since the basic idea behind the time-delay method is quite simple and does not related to any specific models, more reliable constraints can be obtained.

In this paper we present our analysis of limit on $m_{\gamma}$ with a larger GRB radio afterglow sample provided by Chandra \& Frail (2012), thus improving the results of Schaefer (1999). In Section 2 the basic equations and method are presented. Our analysis is based on multi-wavelength radio afterglows as well as prompt emissions for each GRB, and data from more than 60 GRBs are utilized. Our results are presented in Section 3. Also we make an attempt to incorporate synchrotron radiation model for afterglows in our analysis, thus making the time delay shorter, in order to get more stringent constraints. Section 4 summarizes our results and a conclusion is drawn. Throughout our analysis we adopt the standard $\Lambda$CDM cosmology with the cosmological parameters based upon 9-year observations of \textit{Wilkinson Microwave Anisotropy Probe} (WMAP), that is, $H_0 = 69 {\rm{km \, s^{-1} Mpc^{-1}}}$, $\Omega_m = 0.286$, and $\Omega_{\Lambda} = 0.714$ (Hinshaw et al. 2013).

\section{Velocity Dispersion from Non-Zero Photon Mass}
\label{sect:vdnz}

In this section we present our methods of analysis. Supposing the rest mass of photon is $m_\gamma$, according to Einstein's theory of special relativity, the energy of the photon can be expressed as
\begin{equation}\label{}
  E=h\nu=\sqrt{p^2c^2+m_\gamma^2c^4}.
\end{equation}
In the case of  $m_\gamma \neq 0$, the group velocity $v_p$ of photon in vacuum is no longer a constant, and changes with photon energy $E$ (or frequency $\nu$) according to the dispersion relation
\begin{equation}\label{}
  v_p=\frac{\partial{E}}{\partial{p}}=c\sqrt{1-\frac{m_\gamma^2c^4}{E^2}}=c\sqrt{1-A\nu^{-2}}\approx c(1-0.5A\nu^{-2}),
\end{equation}
where $A=\frac{m_{\gamma}^2c^4}{h^2}$.

It is easily seen from Equation (2) that the lower frequency, the slower the photon propagates in vacuum. For explosive events with short time scales such as GRBs, assuming photons with different frequencies emitted simultaneously, the time delay of low energy photons relative to high energy ones thus can be used to calculate the rest mass of a photon. In reality radiations of different bands arise at different times. For example, during a GRB explosion high energy photons should be radiated earlier than X-ray to radio afterglows, and radio afterglows with higher frequencies emerge earlier than lower frequency ones (e. g., see Chandra \& Frail 2012). Therefore, by ignoring such intrinsic time delays, this method can be used to put an upper limit on the photon rest mass.

In our analysis we consider a photon with a higher frequency (energy) $\nu_1$, and another photon with a lower frequency $\nu_2$. Schaefer (1999) did not explicitly clarify what kind of cosmological distance is used to calculate the $\delta t$, while we improve their previous analysis by taking this factor into account. Assuming the high energy photon is emitted at redshift $z$, the comoving distance from source to Earth-based observers for the high energy photon $\nu_1$ with a non-zero mass should be
\begin{equation}\label{}
  D(z,\nu_1)=\frac{c}{H_0}
  \int_0^{z}[1-\frac{1}{2}A\nu_1^{-2}\frac{1}{(1+z')^2}]
  \frac{d{z'}}{\sqrt{\Omega_m(1+z')^3+\Omega_{\Lambda}}}.
\end{equation}
Suppose the low energy photon with $\nu_2$ arrives at Earth later with a nominal redshift difference $z=-\delta z$, $\delta z \ll 1.0$, the comoving distance traveled by this photon is
\begin{equation}\label{}
  D(z,\nu_2)=\frac{c}{H_0}
  \int_{-\delta{z}}^{z}[1-\frac{1}{2}A\nu_2^{-2}\frac{1}{(1+z')^2}]
  \frac{d{z'}}{\sqrt{\Omega_m(1+z')^3+\Omega_{\Lambda}}}.
\end{equation}
Since the two photons should travel the same comoving distance before they reach Earth, we have
\begin{equation}\label{}
  D(z,\nu_1) = D(z,\nu_2).
\end{equation}
Thus the time delay $\delta t$ of the low energy photon $\nu_2$ is
\begin{equation}\label{}
  \delta{t}=\frac{\delta{z}}{H_0}=
  \frac{B}{2H_0}(\nu_2^{-2}-\nu_1^{-2})
  \int_0^{z}\frac{dz'}{(1+z')^2\sqrt{\Omega_m(1+z')^3+\Omega_{\Lambda}}},
\end{equation}
where $B=\frac{2H_0\delta{t}}{(\nu_2^{-2}-\nu_1^{-2})H(z)}$, and $H(z)=\int_0^{z}\frac{{\rm d}z'}{(1+z')^2\sqrt{\Omega_m(1+z')^3+\Omega_{\Lambda}}}$. With a little algebra, the mass of photon can be calculated from the time delay between $\nu_1$ and $\nu_2$ as
\begin{eqnarray}
  m_{\gamma} & = &  B^{1/2} h c^{-2}\nonumber \\
             & = &  \left[ \frac{2H_0\delta t}{\left(\nu_2^{-2}-\nu_1^{-2}\right) H\left(z\right)} \right]^{1/2} h c^{-2} \nonumber\\
             & \simeq & 7.4\times 10^{-48} \left[ \frac{2H_0\delta t}{\left(\nu_2^{-2}-\nu_1^{-2}\right) H\left(z\right)} \right]^{1/2} {\rm{g}}
\end{eqnarray}

It can be seen that in order to get a tighter constraint, a lower $\nu_2$, a larger $z$, as well as a shorter $\delta t$ are prefered. Thus, explosive events with radio emissions occurring at cosmological distances are the most favorable choices. In our analysis we use cosmological GRBs, with their short durations as well as radio afterglows, to put stringent constraints on the upper limit of the photon mass $m_{\gamma}$. However, it should be noted that since $m_{\gamma} \propto \delta t^{1/2}/\left(\nu_2^{-2}-\nu_1^{-2}\right)^{1/2}$, the result of $m_{\gamma}$ is more dependent on $\nu_2$ rather than $\delta t$.

For cosmological GRBs, an intrinsic time delay between prompt gamma-ray emission and radio afterglow does exist, and this should be the most important contributor of the observed $\delta t$. Since prompt emissions are originated from internal interactions of the burst ejecta, and radio afterglows are from later interactions between ejecta and circumburst medium, such a $\delta t_{int}$ is always positive for radio photons. The exact value of $\delta t_{int}$ is hard to know, since early radio afterglows are subject to synchrotron self absorptions, and their starting phases are hard to detect. However, as long as $\delta t_{int} > 0$ stands, what we get from observational data still can provide upper limits of the mass of photons.

Other processes can also be responsible for dispersions between photons with different energies. One of such process is Lorentz invariance violation (LIV), that is, dispersion caused by Planck scale fluctuations of space-time itself. However recent works show that linear LIV does not exist, while time delay from possible higher order items can be completely omitted (e.g., see Abdo et al. 2009, Vasileiou et al. 2013). Besides, high energy photons rather than radio waves suffered most from LIV. Thus we ignore LIV effects in following calculations. Another possible source of time delay is deviation from Einstein's Equivalence Principle (EEP). If such deviation exists, two photons with different energies travel at different speed in gravitational potentials (e.g. see Shapiro 1964, Krauss \& Tremaine 1988, Longo 1988). Currently the best constraints of EEP are $\delta \gamma < 5.9 \times 10^{-14}$ between 1.23 - 1.45 GHz radio photons, and $\delta \gamma < 1.9 \times 10^{-12}$ between MeV and optical (eV) photons, here $\gamma$ is the Parameterized Post-Newtonian (PPN) parameter (Luo et al. 2016). If EEP violation between radio and gamma-ray photons are at the same order of magnitude, this will bring a $\sim$ seconds time delay only, which is much smaller than observed time delay between prompt emission and afterglow of GRBs, and can be neglected. Thus in this work we just assume $\left|\gamma_{MeV} - \gamma_{radio}\right| < 10^{-12}$, and ignore the possible effects of EEP violations. However since no constraint has been performed in such a broad band yet, the possible time delay caused by EEP may bring largest uncertainties here.

Also, plasma effects can change the propagation of radio waves. However, even for cosmological sources within dense medium, such as GRBs, dispersions caused by plasma are in the order of $10^0 \sim 10^2$ seconds only, which is much shorter than the observed time delay between GRB prompt emissions and radio afterglows. And emissions with longer wavelengths should arrive later due to plasma dispersion. Such an effect cannot change our conclusion on upper limits of $m_{\gamma}$. Thus we neglect plasma effects in this paper.

In one word, the observed time delay $\delta t$ between GRB prompt and afterglow emissions came from 3 major sources, non-zero photon mass $\delta t_{\gamma}$, the intrinsic time delay between prompt emission and afterglow $\delta t_{int}$, and plasma effect $\delta t_{plas}$, along with a negligible $\delta t_{others}$ arises from EEP violation, LIV, observational errors, as well as other effects. All the 3 major sources can make lower energy photons arrive later, making $\delta t_{\gamma}>0$, $\delta t_{int}>0$, and $\delta t_{plas}>0$. Thus by considering $\delta t$ comes from non-zero $m_{\gamma}$ completely, what we get is a stringent upper limit of $m_{\gamma}$.

\section{The GRB Sample and the Results}

\label{sect:result}

Chandra \& Frail (2012) compiled radio observations of GRBs between 1997 January and 2011 January, as well as one Fermi burst, GRB 110428A, with a total of 304 GRBs. Based upon this paper, we select our sample for this analysis. First, we put a limit on the mass of photon with the same method as Schaefer (1999), that is, using time delays between high energy prompt detection and the first radio afterglow detection to constrain the mass limit of photon. In our analysis, we choose those GRBs with known redshift $z$ and confirmed radio afterglow detections. Data are selected from various literatures as well as GCN circulars. We compute the time delay $\delta t$ between the trigger time of GRB detectors and the radio observations for each burst, and calculate the upper limit of the photon mass $m_{\gamma}$. For those early afterglow data with large errors, we choose the ones marked as reliable detections in radio light curves in related publications.

In Table 1 we list our data as well as results. Here the frequencies of high energy photon $\nu_1$ are the lower limits of each detector. However, since $\nu_2 \ll \nu_1$ the item $\nu_2^{-2}-\nu_1^{-2}$ can be approximated as $\nu_2^{-2}$, thus in fact the values of $\nu_1$ do not have noticeable effect on our results. All of the redshift $z$ are taken from Chandra \& Frail (2012). We also list the references for our $\delta t$ data in the last column, in which GCN circulars are represented with circular numbers, while published papers are cited by abbreviations, with a complete list presented at the end of the table.

It can be seen from Table 1 that the best result is $m_{\gamma} < 1.062 \times 10^{-44}$ g with GRB 050416A. This GRB has very early radio observations at moderately low radio frequencies ($\sim$ 8.5 GHz). By comparison, The best constraint from Schaefer (1999) is $m_{\gamma} < 4.2 \times 10^{-44}$ g with GRB 980703. Thus we can say that by using early radio afterglow data, our constraints are half orders of magnitude better than those from Schaefer (1999).

However, the main disadvantage of the method of Schaefer (1999) is that, GRB prompt and afterglow photons originate from different regions and different mechanisms, i.e., ejecta interactions for prompt emission and external shocks for afterglows. Although the size difference between prompt and afterglow emission zones is negligible compared with cosmological distance, assuming low and high energy photons radiating from the same region is still too simplified. As a matter of fact, GRB afterglows must arise later than prompt $\gamma$-ray radiations. Thus in order to get more reliable constraints, we also adopt another new method in this analysis, that is, constraining the mass of photon with multi-wavelength radio observations only. In this case, we assume multi-band radio peaks arise at the same time in external shocks despite of afterglow dynamics, and take the time delay between high and low frequency peaks as a result of dispersion induced by non-zero photon mass. It is known that lower frequency radio peaks in GRB afterglows intrinsically appear later than higher frequency ones (e.g., see Wu et al. 2005, as well as Chandra \& Frail 2012), thus from our assumption an upper limit of photon mass can be achieved.

Chandra \& Frail (2012) listed the radio peak flux densities and the peak time for each GRB with radio afterglow light curve observations (i.e, GRBs with more than 3 radio detections in a single radio frequency). Here we analyse the GRBs with (1) radio peak data in two or more bands; (2) flux peaking at higher frequencies earlier than at lower frequencies; and (3) reliable redshift measurements from the whole sample. With all these three criteria we select 23 GRBs with multi-band radio peak data. Also Chandra \& Frail (2012) only used late time data to fit peak time, which means the early reverse shocks have very little effects on fitting results. Thus all peak photons we used here are from forward shock emissions only, which can reduce uncertainties bring out by two-component method used by Schaefer (1999). All of the necessary information of our sample as well as the results derived from multi-band radio peaks are listed in Table 2.

It can be seen that generally speaking, the results derived from the radio peaks are worse than in Table 1. That is because in our second method the values of $\delta t$ are more than 1 order of magnitude larger than those from the first one, thus bring out noticeable effects. One exception is GRB 991208. With a time delay between 1.43 and 8.46 GHz peaks of only $1.1$ days, it can give a constraint of $m_{\gamma} < 1.161 \times 10^{-44}$ g, comparable to (although still slightly larger than) our best result based upon the same method as Schaefer (1999). The major disadvantage of multi-band peak method is that, in Chandra \& Frail (2012), more than half of the peak data are fitting results from forward shock model with uniform circumburst density profile assumption, rather than observations. Therefore, constraints from such radio peaks are more model-dependent than those from the method of Schaefer (1999). The few upper limits drawn from observed peaks are marked with ``*'' in Table 2. These are model-independent results from our second method, and the best one is $m_{\gamma} < 2.804 \times 10^{-44}$ g from GRB 000301C.

In the analysis above, we just assume that the radio peaks of different frequencies were emitted at the same time in a GRB event. However, these peak times can be strongly dependent of dynamical evolution. In the synchrotron radiation model for GRBs, the radio emission peaks at the time when the typical frequency $\nu_m$ drops down to the observational frequency $\nu$, and the frequency $\nu_m$ is  proportional to $t^{-3/2}$ (e.g., see Katz 1994). I.e.,
\begin{equation}\label{}
  \nu_1\propto {t_{p,1}}^{-\frac{3}{2}}, \nu_2\propto {t_{p,2}}^{-\frac{3}{2}},
\end{equation}
Here $t_{p,1}$ and $t_{p,2}$ denote the theoretical values in the framework of the synchrotron radiation model of the peak time at $\nu_1$ and $\nu_2$, respectively. That is
\begin{equation}\label{}
  t_{p,2}=t_{p,1}(\frac{\nu_1}{\nu_2})^{\frac{2}{3}}.
\end{equation}
Assuming $t_{p,1}=t_1$, we can get a time delay $\delta t'$ with synchrotron model correction
\begin{equation}\label{}
  \delta t'=t_2-t_{p,2}=t_2-t_1(\frac{\nu_1}{\nu_2})^{\frac{2}{3}}
\end{equation}
It can be seen that since $\nu_1 > \nu_2$, $\delta t' < \delta t$. Thus in this way we can reduce the value of time delay between $\nu_1$ and $\nu_2$ further for those GRBs with $\delta t' > 0$, and bring out improvement to our results. Our best result based on synchrotron radiation model is $2.346\times 10^{-44}$ g, from GRB 980703, which is still worse than our best constraint without synchrotron corrections. That is because the 1.43 GHz peak of GRB 991208 arrives earlier than the 4.86 GHz peak, making $\delta t' < 0$, thus the data of this burst cannot be used here. Also, the most important issue is that, such a little improvement requires an essential assumption, that is, GRB radio afterglows arise from synchrotron radiation only. Although in most cases it seems to be the case, and in GRB radio afterglows other radiation mechanisms, such as inverse Compton scattering, should not play major roles (e.g., see Wu et al. 2005), such an assumption still may bring extra concerns and uncertainties.

\section{Conclusion and Discussions}
\label{sect:Conclusion}
In this paper, we present our constraints of photon mass $m_{\gamma}$ based upon multi-band GRB data. Using early time radio detections as well as multi-band radio afterglow peaks, we improve the results of Schaefer (1999) by nearly half an order of magnitude. Our best result is $m_{\gamma} < 1.062 \times 10^{-44}$ g from GRB 050416A, from the time delay between early radio afterglow and prompt high energy observations, compared with $4.2 \times 10^{-44}$ g from Schaefer (1999). We also tried to reduce the time difference between multi-band radio peaks by taking synchrotron radiation model into account. However, since the weak dependency between time delay and $m_{\gamma}$, the improvements are not significant, and may bring further uncertainties, therefore such an approach is not favored. We also analyzed several latest GRBs not listed in Chandra \& Frail (2012). However, the results do not show tighter constraints, thus they are not listed in Tables 1 and 2.

Besides possible uncertainties brought by EEP violations, as mentioned in Section 2, other errors from observations and data reductions exist. First of all, early radio observations may suffer from many complicated factors, with synchrotron self absorptions as the most significant one. Synchrotron self absorption can suppress the early fluence of radio afterglow, and can lead to a detection later than the afterglow onset, that is, making the observed $\delta t > \delta t_{\gamma}+\delta t_{int} +\delta t_{plas}+\delta t_{others}$.  Thus these early detections may not truly reflect the onset time of GRBs as well as their afterglows. Also it should be noted that the peak times and fluences of most of samples in Chandra \& Frail (2012) are fitted from multiple observations, rather than taken from direct measurements, thus have considerable errors. However since the dependency between $m_{\gamma}$ and $\delta t$ is relatively weak, that is, $m_{\gamma} \propto \delta t^{1/2}$, even taking all these factors into consideration, the $\delta t$ we use to constrain the photon mass can only vary within 1 order of magnitude in the worst case, and our results still stand.

It should be noted that the upper limit of $m_{\gamma}$ constrained by GRBs may have a best value of $m_{\gamma} \sim 10^{-45}$ g or so. That is because, the radio observations usually cannot start very early after the trigger of GRB detectors. Besides, radio afterglows, especially low frequency afterglows, which are more favorable for $m_{\gamma}$ constraints, suffer strongly from synchrotron self absorption during early times, thus have very low fluxes. However, currently radio afterglow observations are heavily restricted to the sensitivities of instruments, and in most cases only the brightest afterglows at late times can be clearly detected (Chandra \& Frail 2012). Taking GRB 050416A with very early 8.46 GHz radio observations (0.026 day after the GRB trigger) as an example, the upper limit of $m_{\gamma}$ obtained from this GRB is $1.062 \times 10^{-44}$ g. Suppose this burst occuring at a typical redshift $1.8$ rather than the measured one $z=0.650$, and the early radio observations are taken at a lower 1.43 GHz, which is more favorable for $m_{\gamma}$ constraints, we can get $m_{\gamma} < 1.52 \times 10^{-45}$ g. This value might be the best result one can expect from GRB observations, or even all astrophysical explosive events. And with current detector sensitivities, such early low frequency radio observations can be expected only for a few brightest GRBs with ideal circum burst medium conditions. As far as we know, no such early low frequency detection exists so far.

The only exception may be the newly discovered fast radio bursts (FRBs, Lorimer et al. 2007; Thornton et al. 2013). With a duration of $\sim$ ms, and a newly measured redshift $z = 0.492$ for FRB 150418 (Keane et al. 2016), a better constraint can be achieved ($5.2\times 10^{-47}$ g, see Wu et al. 2016). However, currently controversies still remain for FRB redshift measurements, and Williams \& Loeb (2016) noted that the radio ``afterglow'' for FRB 150418 may come from AGN variabilities rather than FRB itself. Thus, constraints from FRBs are with more uncertainties compared to GRBs, and may be not quite reliable. Until the redshift of FRBs can be measured with less uncertainties, constraining the photon mass with GRBs yields the best results among all measurements using time delays.

\section*{Acknowledgments}
This work is partially supported by the National Basic Research Program (``973'' Program)
of China (Grants 2014CB845800 and 2013CB834900), the National Natural Science Foundation
of China (grants Nos. 11322328 and 11433009),
the Youth Innovation Promotion Association (2011231), and the Strategic Priority Research Program
``The Emergence of Cosmological Structures'' (Grant No. XDB09000000) of
the Chinese Academy of Sciences.

%

\begin{table}
\begin{center}
\caption{Constraints of photon mass from time delay between $\gamma$-ray and radio observations.}
\begin{tabular}{lrccccc}

\hline
\hline
GRB &	$\nu_1$ (keV) &	$\nu_2$ (GHz) &	$\delta t$ (days) &	z$^a$ &	$m_{\gamma}$ (g) &	Ref.$^b$ \\
\hline
970508 &	40 &	1.43 &	6.23 &	0.835 &	$2.627 \times 10^{-44}$ &	[1]\\
970508 &	40 &	4.86 &	6.23 &	0.835 &	$8.929 \times 10^{-44}$ &	[1]\\
970508 &	40 &	8.46 &	5.06 &	0.835 &	$1.401 \times 10^{-43}$ &	[1]\\
970828 &	5 &	8.46 &	3.5 &	0.958 &	$1.135 \times 10^{-43}$ &	[2]\\
980329$^c$ &	40 &	4.9 &	9.90 &	2 &	$1.007 \times 10^{-43}$ &	[3]\\
980329 &	40 &	4.9 &	9.90 &	3.9 &	$9.748 \times 10^{-44}$ &	[3]\\
980329 &	40 &	8.3 &	1.07 &	2 &	$5.608 \times 10^{-44}$ &	[3]\\
980329 &	40 &	8.3 &	1.07 &	3.9 &	$5.429 \times 10^{-44}$ &	[3]\\
980425 &	40 &	1.38 &	11.7 &	0.009 &	$2.289 \times 10^{-43}$ &	[4]\\
980425 &	40 &	2.49 &	11.7 &	0.009 &	$4.131 \times 10^{-43}$ &	[4]\\
980425 &	40 &	4.80 &	2.82 &	0.009 &	$3.909 \times 10^{-43}$ &	GCN Circ., 63\\
980425 &	40 &	8.64 &	3.07 &	0.009 &	$7.342 \times 10^{-43}$ &	GCN Circ., 63\\
980703 &	40 &	1.43 &	5.32 &	0.966 &	$2.361 \times 10^{-44}$ &	GCN Circ., 141\\
980703 &	40 &	4.86 &	1.22 &	0.966 &	$3.842 \times 10^{-44}$ &	GCN Circ., 141\\
980703 &	40 &	8.46 &	4.12 &	0.966 &	$1.229 \times 10^{-43}$ &	GCN Circ., 141\\
981226 &	40 &	8.46 &	3.51 &	1.11 &	$1.08 \times 10^{-43}$ &	[5]\\
990123 &	40 &	8.46 &	0.22 &	1.600 &	$2.645 \times 10^{-44}$ &	[6]\\
990506 &	20 &	8.4 &	1.72 &	1.307 &	$7.522 \times 10^{-44}$ &	GCN Circ., 308\\
990510 &	40 &	4.8 &	0.72 &	1.619 &	$2.712 \times 10^{-44}$ &	[7]\\
990510 &	40 &	8.6 &	0.72 &	1.619 &	$4.859 \times 10^{-44}$ &	[7]\\
991208 &	25 &	1.43 &	7.71 &	0.706 &	$3.031 \times 10^{-44}$ &	[8]\\
991208 &	25 &	4.86 &	2.73 &	0.706 &	$6.130 \times 10^{-44}$ &	[8]\\
991208 &	25 &	8.46 &	2.73 &	0.706 &	$1.067 \times 10^{-43}$ &	GCN Circ., 451\\
991208 &	25 &	14.97 &	13.77 &	0.706 &	$4.241 \times 10^{-43}$ &	[8]\\
991208 &	25 &	15 &	3.41 &	0.706 &	$2.115 \times 10^{-43}$ &	GCN Circ., 457\\
991208 &	25 &	22.49 &	13.77 &	0.706 &	$6.371 \times 10^{-43}$ &	[8]\\
991208 &	25 &	30 &	5.44 &	0.706 &	$5.342 \times 10^{-43}$ &	[8]\\
991208 &	25 &	86.24 &	7.43 &	0.706 &	$1.795 \times 10^{-42}$ &	[8]\\
991208 &	25 &	100 &	3.44 &	0.706 &	$1.416 \times 10^{-42}$ &	[8]\\
991208 &	25 &	240 &	3.58 &	0.706 &	$3.467 \times 10^{-42}$ &	GCN Circ., 459\\
991216 &	20 &	4.8 &	1.11 &	1.020 &	$3.586 \times 10^{-44}$ &	GCN Circ., 491\\
991216 &	20 &	8.4 &	1.65 &	1.020 &	$7.650 \times 10^{-44}$ &	GCN Circ., 514\\
991216 &	20 &	8.5 &	1.49 &	1.020 &	$7.536 \times 10^{-44}$ &	GCN Circ., 483\\
991216 &	20 &	15 &	1.14 &	1.020 &	$1.136 \times 10^{-43}$ &	GCN Circ., 489\\
\hline
\end{tabular}
\end{center}
\end{table}

\begin{table}
\begin{center}
\begin{tabular}{lrccccc}

\hline
\hline

GRB &	$\nu_1$ (keV) &	$\nu_2$ (GHz) &	$\delta t$ (days) &	z$^a$ &	$m_{\gamma}$ (g) &	Ref.$^b$ \\
\hline
000301C &	5 &	4.86 &	4.26 &	2.034 &	$6.544 \times 10^{-44}$ &	[9]\\
000301C &	5 &	8.46 &	4.26 &	2.034 &	$1.139 \times 10^{-43}$ &	GCN Circ., 589\\
000301C &	5 &	22.5 &	4.26 &	2.034 &	$3.030 \times 10^{-43}$ &	[9]\\
000301C &	5 &	250 &	2.88 &	2.034 &	$2.768 \times 10^{-42}$ &	GCN Circ., 580\\
000418 &	25 &	4.86 &	12.65 &	1.119 &	$1.207 \times 10^{-43}$ &	[10]\\
000418 &	25 &	8.46 &	10.66 &	1.119 &	$1.929 \times 10^{-43}$ &	GCN Circ., 655\\
000418 &	25 &	15 &	12.32 &	1.119 &	$3.677 \times 10^{-43}$ &	[10]\\
000911 &	25 &	8.46 &	3.06 &	1.059 &	$1.043 \times 10^{-43}$ &	GCN Circ., 795\\
000926 &	25 &	4.86 &	1.98 &	2.039 &	$4.460 \times 10^{-44}$ &	[11]\\
000926 &	25 &	8.46 &	1.17 &	2.039 &	$5.969 \times 10^{-44}$ &	GCN Circ., 805\\
000926 &	25 &	15 &	1.82 &	2.039 &	$1.320 \times 10^{-43}$ &	[11]\\
000926 &	25 &	22.5 &	7.19 &	2.039 &	$3.935 \times 10^{-43}$ &	[11]\\
000926 &	25 &	98.48 &	2.72 &	2.039 &	$1.059 \times 10^{-42}$ &	[11]\\
010222 &	40 &	22 &	0.31 &	1.477 &	$8.238 \times 10^{-44}$ &	GCN Circ., 968\\
010921 &	8 &	4.86 &	25.93 &	0.450 &	$2.130 \times 10^{-43}$ &	GCN Circ., 1107\\
010921 &	8 &	8.46 &	25.93 &	0.450 &	$3.708 \times 10^{-43}$ &	[12]\\
010921 &	8 &	22.5 &	25.93 &	0.450 &	$9.863 \times 10^{-43}$ &	[12]\\
011121 &	40 &	8.64 &	0.88 &	0.362 &	$7.474 \times 10^{-44}$ &	GCN Circ., 1156\\
020405 &	25 &	8.46 &	1.19 &	0.690 &	$7.083 \times 10^{-44}$ &	GCN Circ., 1331\\
020813 &	8 &	8.46 &	1.24 &	1.254 &	$6.469 \times 10^{-44}$ &	GCN Circ., 1490\\
020819 &	8 &	4.86 &	9.72 &	0.410 &	$1.342 \times 10^{-43}$ &	[13]\\
020819 &	8 &	8.46 &	1.75 &	0.410 &	$9.914 \times 10^{-44}$ &	[13]\\
020903 &	2 &	1.5 &	25.7 &	0.250 &	$7.991 \times 10^{-44}$ &	[14]\\
020903 &	2 &	4.9 &	25.7 &	0.250 &	$2.610 \times 10^{-43}$ &	[14]\\
020903 &	2 &	8.46 &	23.8 &	0.250 &	$4.337 \times 10^{-43}$ &	GCN Circ., 1555\\
020903 &	2 &	22.5 &	25.7 &	0.250 &	$1.200 \times 10^{-42}$ &	[14]\\
021004 &	8 &	1.43 &	5.66 &	2.330 &	$2.199 \times 10^{-44}$ &	GCN Circ., 1613\\
021004 &	8 &	4.86 &	5.66 &	2.330 &	$7.472 \times 10^{-44}$ &	GCN Circ., 1613\\
021004 &	8 &	8.46 &	5.66 &	2.330 &	$1.301 \times 10^{-43}$ &	GCN Circ., 1613\\
021004 &	8 &	15 &	1.34 &	2.330 &	$1.122 \times 10^{-43}$ &	GCN Circ., 1588\\
021004 &	8 &	22.5 &	0.78 &	2.330 &	$1.284 \times 10^{-43}$ &	GCN Circ., 1574\\
021004 &	8 &	86.233&	1.42 &	2.330 &	$6.641 \times 10^{-43}$ &	GCN Circ., 1590\\
030115A &	25 &	4.9 &	1.74 &	2.500 &	$4.160 \times 10^{-44}$ &	GCN Circ., 1864\\
030115A &	25 &	8.46 &	2.28 &	2.500 &	$8.221 \times 10^{-44}$ &	GCN Circ., 1827\\

\hline
\end{tabular}
\end{center}
\end{table}

\begin{table}

\begin{center}
\begin{tabular}{lrccccc}

\hline
\hline

GRB &	$\nu_1$ (keV) &	$\nu_2$ (GHz) &	$\delta t$ (days) &	z$^a$ &	$m_{\gamma}$ (g) &	Ref.$^b$ \\
\hline
030329$^d$ &	6 &	1.288 &	2.10 &	0.169 &	$2.287 \times 10^{-44}$ &	GCN Circ., 2073\\
030329 &	6 &	1.43 &	66.53 &	0.169 &	$1.429 \times 10^{-43}$ &	[15]\\
030329 &	6 &	4.86 &	1.05 &	0.169 &	$6.103 \times 10^{-44}$ &	[15]\\
030329 &	6 &	8.46 &	0.58 &	0.169 &	$7.896 \times 10^{-44}$ &	GCN Circ., 2014\\
030329 &	6 &	15.2 &	1.21 &	0.169 &	$2.049 \times 10^{-43}$ &	GCN Circ., 2043\\
030329 &	6 &	22.5 &	2.65 &	0.169 &	$4.489 \times 10^{-43}$ &	[15]\\
030329 &	6 &	23 &	4.98 &	0.169 &	$6.290 \times 10^{-43}$ &	GCN Circ., 2089\\
030329 &	6 &	43.3 &	2.65 &	0.169 &	$8.638 \times 10^{-43}$ &	[15]\\
030329 &	6 &	90 &	4.98 &	0.169 &	$2.461 \times 10^{-42}$ &	GCN Circ., 2089\\
030329 &	6 &	350 &	4.92 &	0.169 &	$9.514 \times 10^{-42}$ &	GCN Circ., 2088\\
031203 &	20 &	1.43 &	39.37 &	0.105 &	$1.347 \times 10^{-43}$ &	[16]\\
031203 &	20 &	4.86 &	4.43 &	0.105 &	$1.536 \times 10^{-43}$ &	[16]\\
031203 &	20 &	8.46 &	1.60 &	0.105 &	$1.606 \times 10^{-43}$ &	[16]\\
031203 &	20 &	22.5 &	13.46 &	0.105 &	$1.239 \times 10^{-42}$ &	[16]\\
050315 &	15 &	8.5 &	0.80 &	1.950 &	$4.976 \times 10^{-44}$ &	GCN Circ., 3102\\
050401 &	15 &	8.5 &	5.69 &	2.898 &	$1.295 \times 10^{-43}$ &	GCN Circ., 3187\\
050416A &	15 &	1.43 &	76.46 &	0.650 &	$9.736 \times 10^{-44}$ &	[17]\\
050416A &	15 &	4.86 &	5.58 &	0.650 &	$8.938 \times 10^{-44}$ &	GCN Circ., 3318\\
050416A$^e$ &	15 &	8.46 &	0.026 &	0.650 &	$1.062 \times 10^{-44}$ &	[17]\\
050525A &	15 &	22.5 &	0.42 &	0.606 &	$1.155 \times 10^{-43}$ &	GCN Circ., 3495\\
050603 &	15 &	8.5 &	0.35 &	2.821 &	$3.216 \times 10^{-44}$ &	GCN Circ., 3513\\
050724 &	15 &	8.5 &	0.57 &	0.258 &	$6.665 \times 10^{-44}$ &	GCN Circ., 3676\\
050730$^f$ &	15 &	4.9 &	5.63 &	3.968 &	$7.348 \times 10^{-44}$ &	GCN Circ., 3781\\
050730 &	15 &	8.5 &	2.20 &	3.968 &	$7.968 \times 10^{-44}$ &	GCN Circ., 3761\\
050820A &	15 &	4.86 &	2.15 &	2.615 &	$4.575 \times 10^{-44}$ &	[18]\\
050820A &	15 &	8.46 &	0.116 &	2.615 &	$1.850 \times 10^{-44}$ &	[18]\\
050904 &	15 &	8.46 &	34.18 &	6.290 &	$3.101 \times 10^{-43}$ &	[19]\\
051022 &	6 &	4.9 &	1.09 &	0.809 &	$3.790 \times 10^{-44}$ &	GCN Circ., 4158\\
051022 &	6 &	8.5 &	1.54 &	0.809 &	$7.815 \times 10^{-44}$ &	GCN Circ., 4154\\
051022 &	6 &	100 &	1.35 &	0.809 &	$8.609 \times 10^{-43}$ &	GCN Circ., 4157\\
051109A &	15 &	8.5 &	2.10 &	2.346 &	$7.957 \times 10^{-44}$ &	GCN Circ., 4244\\
051221A &	15 &	8.5 &	0.91 &	0.547 &	$6.601 \times 10^{-44}$ &	GCN Circ., 4416\\
060218 &	15 &	8.46 &	1.85 &	0.033 &	$2.956 \times 10^{-43}$ &	GCN Circ., 4794\\
061121 &	15 &	8.46 &	0.74 &	1.315 &	$4.965 \times 10^{-44}$	&	GCN Circ., 5843\\
\hline
\end{tabular}
\end{center}
\end{table}

\begin{table}

\begin{center}
\begin{tabular}{lrccccc}

\hline
\hline

GRB &	$\nu_1$ (keV) &	$\nu_2$ (GHz) &	$\delta t$ (days) &	z$^a$ &	$m_{\gamma}$ (g) &	Ref.$^b$ \\
\hline

061222A &	15 &	8.46 &	0.86 &	2.088 &	$5.108 \times 10^{-44}$ &	GCN Circ., 5997\\
070125 &	30 &	4.9 &	5.39 &	1.548 &	$7.610 \times 10^{-44}$ &	GCN Circ., 6063\\
070125 &	30 &	8.46 &	4.01 &	1.548 &	$1.133 \times 10^{-43}$ &	[20]\\
070125 &	30 &	14.94 &	13.07 &	1.548 &	$3.380 \times 10^{-43}$ &	[20]\\
070125 &	30 &	22.5 &	10.73 &	1.548 &	$4.930 \times 10^{-43}$ &	[20]\\
070125 &	30 &	95 &	10.98 &	1.548 &	$2.106 \times 10^{-42}$ &	[20]\\
070125 &	30 &	250 &	6.52 &	1.548 &	$4.270 \times 10^{-42}$ &	[20]\\
070612A &	15 &	4.88 &	18.75 &	0.617 &	$1.667 \times 10^{-43}$ &	GCN Circ., 6600\\
070612A$^g$ &	15 &	4.9 &	3.23 &	0.617 &	$6.946 \times 10^{-44}$ &	GCN Circ., 6549\\
071003 &	15 &	4.86 &	3.84 &	1.604 &	$6.347 \times 10^{-44}$ &	[21]\\
071003 &	15 &	8.46 &	1.76 &	1.604 &	$7.480 \times 10^{-44}$ &	GCN Circ., 6853\\
071020 &	15 &	8.46 &	2.17 &	2.146 &	$8.098 \times 10^{-44}$ &	GCN Circ., 6978\\
071122 &	15 &	8.46 &	2.88 &	1.140 &	$9.997 \times 10^{-44}$ &	GCN Circ., 7132\\
080319B &	15 &	4.86 &	2.30 &	0.937 &	$5.305 \times 10^{-44}$ &	GCN Circ., 7506\\
080319B$^h$ &	15 &	4.9 &	3.58 &	0.937 &	$6.673 \times 10^{-44}$ &	GCN Circ., 7507\\
080603A$^i$ &	20 &	8.46 &	1.92 &	1.687 &	$7.773 \times 10^{-44}$ &	GCN Circ., 7843\\
080603A &	20 &	8.46 &	3.95 &	1.687 &	$1.115 \times 10^{-43}$ &	GCN Circ., 7855\\
080810 &	15 &	8.46 &	2.81 &	3.350 &	$9.008 \times 10^{-44}$ &	GCN Circ., 8103\\
090313 &	15 &	4.9 &	7.41 &	3.375 &	$8.471 \times 10^{-44}$ &	GCN Circ., 9016\\
090313 &	15 &	8.46 &	5.85 &	3.375 &	$1.299 \times 10^{-43}$ &	GCN Circ., 9011\\
090313 &	15 &	14.5 &	2.70 &	3.375 &	$1.513 \times 10^{-43}$ &	GCN Circ., 9003\\
090313 &	15 &	92.5 &	1.02 &	3.375 &	$5.933 \times 10^{-43}$ &	GCN Circ., 9005\\
090323 &	8 &	4.9 &	4.74 &	3.57 &	$6.762 \times 10^{-44}$ &	GCN Circ., 9047\\
090323 &	8 &	8.46 &	4.38 &	3.57 &	$1.122 \times 10^{-43}$ &	GCN Circ., 9043\\
090328 &	8 &	8.46 &	2.59 &	0.736 &	$1.030 \times 10^{-43}$ &	GCN Circ., 9060\\
090418 &	15 &	8.46 &	1.00 &	1.608 &	$5.637 \times 10^{-44}$ &	GCN Circ., 9166\\
090423 &	15 &	8.46 &	7.72 &	8.260 &	$1.471 \times 10^{-43}$ &	[22]\\
090424 &	15 &	8.46 &	1.50 &	0.544 &	$8.447 \times 10^{-44}$ &	GCN Circ., 9260\\
090618 &	15 &	4.9 &	1.43 &	0.540 &	$4.787 \times 10^{-44}$ &	GCN Circ., 9538\\
090618 &	15 &	8.46 &	0.98 &	0.540 &	$6.842 \times 10^{-44}$ &	GCN Circ., 9533\\
090618 &	15 &	14.6 &	0.76 &	0.540 &	$1.040 \times 10^{-43}$ &	GCN Circ., 9532\\
090715B &	15 &	8.46 &	4.26 &	3.000 &	$1.114 \times 10^{-43}$ &	GCN Circ., 9695\\
090902B &	8 &	4.8 &	1.13 &	1.883 &	$3.349 \times 10^{-44}$ &	GCN Circ., 9883\\
090902B &	8 &	8.46 &	1.54 &	1.883 &	$6.892 \times 10^{-44}$ &	GCN Circ., 9889\\
091020 &	15 &	8.46 &	0.76 &	1.710 &	$4.884 \times 10^{-44}$ &	GCN Circ., 10088\\
100414A &	8 &	4.8 &	13.57 &	1.368 &	$1.200 \times 10^{-43}$ &	GCN Circ., 10697\\
100414A &	8 &	8.46 &	13.08 &	1.368 &	$2.077 \times 10^{-43}$ &	GCN Circ., 10698\\
\hline
\end{tabular}
\end{center}
\end{table}

\begin{table}

\begin{center}
\begin{tabular}{lrccccc}

\hline
\hline

GRB &	$\nu_1$ (keV) &	$\nu_2$ (GHz) &	$\delta t$ (days) &	z$^a$ &	$m_{\gamma}$ (g) &	Ref.$^b$ \\
\hline
100418A &	15 &	4.8 &	1.97 &	0.620 &	$5.307 \times 10^{-44}$ &	GCN Circ., 10647\\
100418A &	15 &	5.5 &	2 &	0.620 &	$6.127 \times 10^{-44}$ &	[23]\\
100418A &	15 &	8.46 &	3 &	0.620 &	$1.154 \times 10^{-43}$ &	[23]\\
100418A &	15 &	9 &	2 &	0.620 &	$1.003 \times 10^{-43}$ &	[23]\\
100418A &	15 &	345 &	0.66 &	0.620 &	$2.208 \times 10^{-42}$ &	GCN Circ., 10630\\
100901A &	15 &	4.5 &	4.90 &	1.408 &	$6.737 \times 10^{-44}$ &	GCN Circ., 11257\\
100901A$^j$ &	15 &	4.9 &	2.28 &	1.408 &	$5.004 \times 10^{-44}$ &	GCN Circ., 11221\\
100901A &	15 &	7.9 &	4.90 &	1.408 &	$1.183 \times 10^{-43}$ &	GCN Circ., 11257\\
100906A &	15 &	13.5 &	2.34 &	1.727 &	$1.366 \times 10^{-43}$ &	GCN Circ., 11259\\
\hline
\end{tabular}
\end{center}

Notes:\\
$^a$ All redshift data are adopted from Chandra \& Frail (2012)\\
$^b$ Abbreviations for the references are as follows: [1] Frail et al. (1999a); [2] Djorgovski et al. (2001); [3] Taylor et al. (1998); [4] Kulkarni et al. (1998); [5] Frail et al. (1999b); [6] Kulkarni et al. (1999); [7] Harrison et al. (1999); [8] Galama et al. (2000); [9] Berger et al. (2000); [10] Berger et al. (2001); [11] Harrison et al. (2001); [12] Price et al. (2002); [13] Jakobsson et al. (2005); [14] Soderberg et al. (2004a); [15] Berger et al. (2003); [16] Soderberg et al. (2004b); [17] Soderberg et al. (2007); [18] Cenko et al. (2006); [19] Frail et al. (2006); [20] Chandra et al. (2008); [21] Perley et al. (2008); [22] Chandra et al. (2010); [23] Moin et al. (2013).\\
$^c$ Chandra \& Frail (2012) presents a redshift range ($z \sim 2-3.9 $) for this burst. Thus we calculate the $m_{\gamma}$ with upper and lower limits of $z$, respectively.\\
$^d$ Possible detection. \\
$^e$ The 0.026 day observation has very large error bars ($20 \pm 51$ $\mu$Jy), as shown in Soderberg et al. (2007). However in Fig. 6 of Chandra \& Frail (2012) this data point is marked as ``radio detected''. Thus in our analysis we take this early observation into consideration.\\
$^f$ Possible detection. \\
$^g$ Possible detection. \\
$^h$ Possible detection.\\
$^i$ Possible detection.\\
$^j$ Possible detection.\\

\end{table}

\newpage

\begin{table}

\caption{Constraints of photon mass from multi-band radio afterglow peaks.}

\begin{center}
\begin{tabular}{l c c c c c c}
\hline
\hline
GRB & $\nu_1$(GHz) & $\nu_2$(GHz)&$t_1$(day) & $t_2$(day) & z & m$_\gamma$ (g)  \\
\hline
970508	&4.86&	1.43&	57.6&	179.1&	0.835&	$1.214 \times 10^{-43}$ \\
970508	&8.46&	1.43&	37.2&	179.1&	0.835&	$1.272 \times 10^{-43}$ \\
970508	&8.46&	4.86&	37.2&	57.6&	0.835&	$1.974 \times 10^{-43}$ \\
980425	&2.5&	1.38&	32.7&	47.1&	0.009&	$3.046 \times 10^{-43}$ \\
980425	&4.8&	1.38&	18.3&	47.1&	0.009&	$3.750 \times 10^{-43}$ \\
980425	&4.8&	2.5&	18.3&	32.7&	0.009&	$5.390 \times 10^{-43}$ \\
980425	&8.64&	1.38&	12.7&	47.1&	0.009&	$3.977 \times 10^{-43}$ \\
980425	&8.64&	2.5&	12.7&	32.7&	0.009&	$5.665 \times 10^{-43}$ \\
980425	&8.64&	4.8&	12.7&	18.3&	0.009&	$6.690 \times 10^{-43}$ \\
980703	&4.86&	1.43&	9.1&	25.4&	0.966&	$4.324 \times 10^{-44}$ \\
980703	&8.46&	1.43&	10&	25.4&	0.966&	$4.076 \times 10^{-44}$ \\
990510	&8.46&	4.86&	4.2&	9.2&	1.619&	$8.840 \times 10^{-44}$ \\
991208	&8.46&	1.43&	7.8&	8.9&	0.706&	$1.161 \times 10^{-44}$ \\
991208	&8.46&	4.86&	7.8&	12.8&	0.706&	$1.014 \times 10^{-43}$ \\
991208	&15&	1.43&	5.4&	8.9&	0.706&	$2.052 \times 10^{-44}$ \\
991208	&15&	4.86&	5.4&	12.8&	0.706&	$1.067 \times 10^{-43}$ \\
991208	&15&	8.46&	5.4&	7.8&	0.706&	$1.212 \times 10^{-43}$ \\
991216*	&15&	4.86&	1.3&	17.4&	1.020&	$1.461 \times 10^{-43}$ \\
000301C*	&15&	4.86&	3.6&	4.3&	2.034&	$2.804 \times 10^{-44}$ \\
000301C*	&15&	8.46&	3.6&	14.1&	2.034&	$2.166 \times 10^{-43}$ \\
000301C	&22.5&	8.46&	4.3&	14.1&	2.034&	$1.884 \times 10^{-43}$ \\
000418	&8.46&	4.86&	18.1&	27&	1.119&	$1.237 \times 10^{-43}$ \\
000418	&15&	4.86&	12.3&	27&	1.119&	$1.375 \times 10^{-43}$ \\
000418	&15&	8.46&	12.3&	18.1&	1.119&	$1.723 \times 10^{-43}$ \\
000418	&22.5&	4.86&	14.6&	27&	1.119&	$1.224 \times 10^{-43}$ \\
000418	&22.5&	8.46&	14.6&	18.1&	1.119&	$1.193 \times 10^{-43}$ \\
000911	&8.46&	4.86&	3.1&	11.1&	1.059&	$1.183 \times 10^{-43}$ \\
000926	&8.46&	4.86&	12.1&	16.9&	2.039&	$8.485 \times 10^{-44}$ \\
000926	&15&	4.86&	4.7&	16.9&	2.039&	$1.170 \times 10^{-43}$ \\
000926	&15&	8.46&	4.7&	12.1&	2.039&	$1.818 \times 10^{-43}$ \\
000926	&22.5&	4.86&	14.6&	16.9&	2.039&	$4.924 \times 10^{-44}$ \\
000926	&22.5&	8.46&	7.2&	12.1&	2.039&	$1.318 \times 10^{-43}$ \\
\hline
\end{tabular}
\end{center}

\end{table}

\begin{table}

\begin{center}
\begin{tabular}{l c c c c c c}
\hline
\hline
GRB & $\nu_1$(GHz) & $\nu_2$(GHz)&$t_1$(day) & $t_2$(day) & z & m$_\gamma$ (g)  \\
\hline
010921	&8.46&	4.86&	27&	31.5&	0.45&	$1.084 \times 10^{-43}$ \\
010921	&22.5&	4.86&	25.9&	31.5&	0.45&	$1.014 \times 10^{-43}$ \\
010921*	&22.5&	8.46&	25.9&	27&	0.45&	$8.243 \times 10{-44}$ \\
020903	&4.86&	1.43&	36.7&	91&	0.25&	$1.159 \times 10^{-43}$ \\
020903*	&8.46&	1.43&	23.8&	91&	0.25&	$1.250 \times 10^{-43}$ \\
020903	&8.46&	4.86&	23.8&	36.7&	0.25&	$2.241 \times 10^{-43}$ \\
021004	&8.46&	4.86&	18.7&	32.2&	2.33&	$1.410 \times 10^{-43}$ \\
021004	&15&	4.86&	4.1&	32.2&	2.33&	$1.760 \times 10^{-43}$ \\
021004	&15&	8.46&	4.1&	18.7&	2.33&	$2.530 \times 10^{-43}$ \\
021004	&22.5&	4.86&	8.7&	32.2&	2.33&	$1.559 \times 10^{-43}$ \\
021004	&22.5&	8.46&	8.7&	18.7&	2.33&	$1.866 \times 10^{-43}$ \\
030329	&4.86&	1.43&	32.9&	78.6&	0.169&	$1.240 \times 10^{-43}$ \\
030329	&8.46&	1.43&	17.3&	78.6&	0.169&	$1.392 \times 10^{-43}$ \\
030329	&8.46&	4.86&	17.3&	32.9&	0.169&	$2.874 \times 10^{-43}$ \\
030329	&15&	1.43&	10.9&	78.6&	0.169&	$1.448 \times 10^{-43}$ \\
030329	&15&	4.86&	10.9&	32.9&	0.169&	$2.953 \times 10^{-43}$ \\
030329	&15&	8.46&	10.9&	17.3&	0.169&	$3.176 \times 10^{-43}$ \\
030329	&22.5&	1.43&	8.4&	78.6&	0.169&	$1.471 \times 10^{-43}$ \\
030329	&22.5&	4.86&	8.4&	32.9&	0.169&	$3.019 \times 10^{-43}$ \\
030329	&22.5&	8.46&	8.4&	17.3&	0.169&	$3.338 \times 10^{-43}$ \\
030329	&22.5&	15&	8.4&	10.9&	0.169&	$3.899 \times 10^{-43}$ \\
030329	&43&	1.43&	5.8&	78.6&	0.169&	$1.496 \times 10^{-43}$ \\
030329	&43&	4.86&	5.8&	32.9&	0.169&	$3.120 \times 10^{-43}$ \\
030329	&43&	8.46&	5.8&	17.3&	0.169&	$3.586 \times 10^{-43}$ \\
030329	&43&	15&	5.8&	10.9&	0.169&	$4.430 \times 10^{-43}$ \\
030329	&43&	22.5&	5.8&	8.4&	0.169&	$5.217 \times 10^{-43}$ \\
031203	&4.86&	1.43&	58.4&	65.5&	0.105&	$5.986 \times 10^{-44}$ \\
031203	&8.46&	1.43&	48&	65.5&	0.105&	$9.112 \times 10^{-44}$ \\
031203	&8.46&	4.86&	48&	58.4&	0.105&	$2.875 \times 10^{-43}$ \\
031203	&22.5&	1.43&	13.5&	65.5&	0.105&	$1.551 \times 10^{-43}$ \\
031203	&22.5&	4.86&	13.5&	58.4&	0.105&	$5.007 \times 10^{-43}$ \\
031203	&22.5&	8.46&	13.5&	48&	0.105&	$8.051 \times 10^{-43}$ \\
051022	&8.46&	4.86&	5.2&	57&	0.809&	$3.166 \times 10^{-43}$ \\
\hline
\end{tabular}
\end{center}

\end{table}

\begin{table}

\begin{center}
\begin{tabular}{l c c c c c c}
\hline
\hline
GRB & $\nu_1$(GHz) & $\nu_2$(GHz)&$t_1$(day) & $t_2$(day) & z & m$_\gamma$ (g)  \\
\hline
060218	&4.86&	1.43&	3.8&	4.9&	0.033&	$4.032 \times 10^{-44}$ \\
060218	&8.46&	1.43&	2&	4.9&	0.033&	$6.348 \times 10^{-44}$ \\
060218	&8.46&	4.86&	2&	3.8&	0.033&	$2.047 \times 10^{-43}$ \\
060218*	&22.5&	1.43&	3&	4.9&	0.033&	$5.075 \times 10^{-44}$ \\
060218	&22.5&	4.86&	3&	3.8&	0.033&	$1.144 \times 10^{-43}$ \\
070125	&22.5&	15&	13.6&	18.6&	1.548&	$3.010 \times 10^{-43}$ \\
070612A	&8.46&	4.86&	84.1&	140.3&	0.617&	$3.511 \times 10^{-43}$ \\
071010B	&8.46&	4.86&	4.2&	12.5&	0.947&	$1.229 \times 10^{-43}$ \\
100414A	&8.46&	4.86&	8&	38&	1.368&	$2.207 \times 10^{-43}$ \\
100418A	&8.46&	4.86&	47.6&	70.3&	0.62&	$2.228 \times 10^{-43}$ \\
100814A	&7.9&	4.5&	10.4&	13&	1.44&	$5.955 \times 10^{-44}$ \\
\hline
\end{tabular}
\end{center}
Notes: Peak and redshift data are adopted from Chandra \& Frail (2012).\\
*: Constraints with observed (rather than fitted) radio peaks.
\end{table}

\end{document}